\documentclass[doublecol]{epl2}
\usepackage[T1]{fontenc}
\usepackage[latin1]{inputenc}
\usepackage{amsmath}
\usepackage{graphicx}

\title{Aharonov-Bohm effect in undoped graphene: Magnetotransport via evanescent waves}

\shorttitle{Aharonov-Bohm effect in graphene}

\author{M. I. Katsnelson}
\shortauthor{M. I. Katsnelson}

\institute{Institute for Molecules and Materials, Radboud
University Nijmegen, Heyendaalseweg 135, 6525 AJ Nijmegen, The
Netherlands
 }

\pacs{72.80.Vp}{Electronic transport in graphene}

\pacs{73.23.Ad}{Ballistic transport}

\pacs{03.65.Pm}{Relativistic wave equations}

\abstract{Using conformal mapping technique, compact and general
analytic expressions for the effects of magnetic fluxes on
conductance and Fano factor of undoped graphene nanoflakes in
pseudodiffusive regime are derived.}

\begin{document}

\maketitle

Massless Dirac fermions in two dimensions have amazing property,
the finite conductivity (of order of conductance quantum $e^2/h$)
in the limit of zero charge carrier concentration and no disorder
\cite{Letal94}. It can be understood in terms of transport via
evanescent waves (zero modes of the Dirac operator) \cite{K06}.
Despite we deal with an ideal situation (no scattering by defects)
some transport properties, such as statistics of shot noise, are
reminiscent of those for disordered metals \cite{Tetal06} which
justifies a frequently used term ``pseudodiffusive regime''. There
are many theoretical \cite{K06,Tetal06,Petal07,KG08,RRW09,SOGM09}
and experimental \cite{Metal07,Detal08} works studying this regime
in a context of graphene. This situation is very special. For
conventional electron gas in semiconductors, in the absence of
disorder, the states with definite energy (eigenstates of the
Hamiltonian) can be simultaneously the states with definite
current (eigenstates of the current operator) and it is the
disorder that results in non-conservation of the current and
finite conductivity. Contrary, for Dirac fermions the current
operator does not commute with the Hamiltonian
(``Zitterbewegung'') which can be considered as a kind of
intrinsic disorder \cite{K06,AK07}. More detailed understanding of
the pseudodiffusive transport in graphene is therefore of a
general interest for quantum statistical physics and physical
kinetics.

Here we discuss peculiarities of the Aharonov-Bohm (AB) effect
\cite {AB59,OP85} in the pseudodiffusive regime for graphene. The
AB effect in graphene has been studied already theoretically \cite
{Retal07,Jetal09,Wetal09} and experimentally
\cite{Retal08,Metal09}. Combining a general consideration of
zero-energy states for massless Dirac fermions \cite{AC79} with
recent elegant treatment of the transport via evanescent waves by
conformal transformation technique \cite{RRW09} we derive simple
and general rigorous formulas for any graphene flake topologically
equivalent to the ring, avoiding both numerical simulations and
explicit solutions of Shr\"{o}dinger equation for some particular
cases.

For the case of zero energy $E=0$ (undoped graphene) the Dirac
equation
\begin{equation}
\mathbf{\sigma }\left( -i\nabla \mathbf{-A}\right) \psi =E\psi
\end{equation}
for the two-component spinor $\psi ,$ $\mathbf{A}$ is the vector
potential and we use the units $\hbar =e=1,$ splits into two
independent equations for the components $\psi _\sigma $ ($\sigma
=\pm $):
\begin{equation}
\left( -i\frac \partial {\partial x}+\sigma \frac \partial
{\partial y}-A_x-i\sigma A_y\right) \psi _\sigma =0  \label{eq}
\end{equation}
We will use the gauge $\nabla \mathbf{A}=0$ and introduce scalar
magnetic potential $\varphi $ by
\begin{equation}
A_x=\frac{\partial \varphi }{\partial y},A_y=-\frac{\partial \varphi }{%
\partial x},
\end{equation}
thus,
\begin{equation}
\nabla ^2\varphi =-B  \label{laplace}
\end{equation}
where $B$ is the magnetic induction. The vector potential can be
eliminated by a substitution \cite{AC79}
\begin{equation}
\psi _\sigma =e^{\sigma \varphi }\chi _\sigma ,  \label{sub}
\end{equation}
Eq.(\ref{eq}) being satisfied for an arbitrary analytic function
$\chi _{+}\left( z=x+iy\right) $ and complex conjugated analytic
function $\chi _{\_}\left( \overline{z}\right) .$ The latter can
be found from boundary conditions. To consider the pseudodiffusive
transport one can assume that the leads 1 and 2 attached to the
undoped graphene flake are made from heavily doped graphene, such
that the Fermi wavelength of electrons  within the leads is much
smaller than any geometric size of the system under consideration.
Thus, the boundary conditions read\cite{K06,Tetal06,KG08}
\begin{eqnarray}
1+r &=&\psi _{+}^{(1)}  \nonumber  \label{boun} \\
1-r &=&\psi _{-}^{(1)}  \nonumber \\
t &=&\psi _{+}^{(2)}  \nonumber \\
t &=&\psi _{-}^{(2)}  \label{boun} \\
&&  \nonumber
\end{eqnarray}
where $r$ and $t$ are reflection and transmission coefficients,
respectively, and superscripts 1 and 2 label the boundaries
attached to the corresponding leads. If the boundary of the sample
is simply connected  one can always choose $\varphi =0$ at the
boundary and, thus, the magnetic field disappears from
Eq.(\ref{boun}); this fact was used in Ref.\cite{SOGM09} as a very
elegant way to prove that a random vector potential does not
effect on the value of minimal conductivity. Further we will
consider a sample with a topology of the ring where the scalar
potential $\varphi $ is still constant at each boundary but these
constants, $\varphi _1$ and $\varphi _2$ are different. Also, by
symmetry,
\begin{equation}
\chi _{+}^{(2)}/\chi _{+}^{(1)}=\chi _{-}^{(1)}/\chi _{-}^{(2)}
\label{sym}
\end{equation}
Further derivation follows Ref. \cite{RRW09} and the answer for
the transmission coefficient $T=\left| t\right| ^2$ has the form:
\begin{equation}
T_j=\frac 1{\cosh ^2\left[ 2\left( j+a\right) \ln \Lambda \right]
} \label{T}
\end{equation}
where $j=\pm 1/2,\pm 3/2,...$ labels zero modes of the Dirac equation, $%
\Lambda $ is determined by a conformal transformation of our flake
to the rectangle and
\begin{equation}
a=\frac{\varphi _2-\varphi _1}{2\ln \Lambda }  \label{a}
\end{equation}
For simplicity, we will consider further the case of the Corbino
disc with inner radius $R_1$ and outer radius $R_2$, when
\cite{RRW09}
\begin{equation}
\Lambda =\sqrt{R_2/R_1}  \label{lambda}
\end{equation}
The conductance $G$ (per spin per valley) and Fano factor of the shot noise $%
F$ are expressed via the transmission coefficients (\ref{T}) as
\begin{eqnarray}
G &=&\frac{e^2}h\sum_{j=-\infty }^\infty T_j,  \nonumber \\
F &=&1-\frac{\sum_{j=-\infty }^\infty T_j^2}{\sum_{j=-\infty
}^\infty T_j} \label{land}
\end{eqnarray}
To calculate the sums in Eq.(\ref{land}) one can use the Poisson
summation formula
\begin{equation}
\sum\limits_{n=-\infty }^\infty f\left( n+x\right)
=\sum\limits_{k=-\infty }^\infty \exp \left[ -2i\pi kx\right]
\widetilde{f}_k
\end{equation}
where $n,k$ are all integers and $\widetilde{f}_k$ is the Fourier
transform of the function $f\left( x\right) .$ Substituting
Eq.(\ref{T}) into (\ref{land}) one finds a compact and general
answer for the effect of magnetic field on the transport
characteristics:
\begin{equation}
G=\frac{e^2}{h\ln \Lambda }\left[ 1+2\sum_{k=1}^\infty \left(
-1\right) ^k\cos \left( 2\pi ka\right) \alpha_k \right]
\label{answ1}
\end{equation}
\begin{equation}
F=1-\frac 23\left[ \frac{1+2\sum_{k=1}^\infty \left( -1\right)
^k\cos \left( 2\pi ka\right) \alpha_k\left( 1+\frac{\pi ^2k^2}{4\ln ^2\Lambda }\right) }{%
1+2\sum_{k=1}^\infty \left( -1\right) ^k\cos \left( 2\pi ka\right)
\alpha_k}\right] \label{answ2}
\end{equation}
where
\begin{equation}
\alpha_k = \frac{\pi ^2k/2\ln \Lambda }{\sinh \left( \pi ^2k/2\ln
\Lambda \right) } \label{alpha}
\end{equation}

Eq.(\ref{laplace}) can be solved explicitly for radially symmetric
distributions of the magnetic field $B\left( r\right) $:
\begin{equation}
\varphi _2-\varphi _1=\frac \Phi {2\pi }\ln \left(
\frac{R_2}{R_1}\right)
+\int\limits_{R_1}^{R_2}\frac{dr}r\int\limits_0^rdr^{\prime
}r^{\prime }B\left( r^{\prime }\right)
\end{equation}
where $\Phi $ is the magnetic flux through the inner ring. In the
case of AB effect where all magnetic flux is concentrated within
the inner ring one has
\begin{equation}
a=\frac{e\Phi }{2\pi \hbar c}  \label{aa}
\end{equation}
where we have restored world constants.

Due to the large factor $\pi^2$ in the argument of sinh in
Eq.(\ref{alpha}) only the terms with $k=1$ should be kept in
Eqs.(\ref{answ1}) and (\ref{answ2}) for all realistic shapes,
thus,
\begin{equation}
G=G_0\left[ 1- \frac{4\pi ^2}{\ln \left( R_2/R_1\right)} \exp
\left( -\frac{\pi ^2}{\ln \left( R_2/R_1\right)} \right) \cos
\left( \frac{e\Phi }{\hbar c}\right)\right] \label{answ3}
\end{equation}
\begin{equation}
F=\frac{1}{3}+\frac{8\pi^4}{3\ln^3\left( R_2/R_1 \right)}
\exp\left(- \frac{\pi^2}{\ln \left( R_2/R_1 \right)} \right) \cos
\left( \frac{e\Phi}{\hbar c} \right)  \label{answ4}
\end{equation}
where $G_0=\frac{2e^2}{h\ln(R_2/R_1)}$.

Oscillating contributions to $G$ and $F$ are exponentially small
for very thin rings but, for sure, measurable if the ring is thick
enough. For $R_2/R_1=5$ their amplitudes are 5.3\% and 40\%,
respectively.

Pseudomagnetic fields describing by the vector potential can be
created by deformations of the graphene flake \cite{FGM08,GKG09}.
Expressions (\ref {answ1}), (\ref{answ2}) can be applied also in
this situation.

Consider now a generic case with the magnetic (or pseudomagnetic \cite{FGM08}%
) field $B=0$ within the flake. Then, the solution of
Eq.(\ref{laplace}) can be obtained from the solution for the disc
by the same conformal transformation which determines the function
$\Lambda $. One can see immediately that Eq.(\ref{aa}) remains the
same. As for the expressions (\ref{answ3}), (\ref{answ4})  they
can be rewritten in terms of experimentally measurable quantity
$G_0,$
\begin{equation}
G=G_0\left[ 1-\frac{4\pi ^2}{\beta} \exp \left( -\pi ^2/\beta
\right) \cos \left( \frac{e\Phi }{\hbar c}\right)\right]
\label{fine1}
\end{equation}
\begin{equation}
F=\frac 13+ \frac{8\pi ^4}{3\beta ^3}\exp \left( -\pi ^2/\beta
\right) \cos \left( \frac{e\Phi }{\hbar c}\right) \label{fine2}
\end{equation}
where $\beta =2e^2/hG_0$ and we assume $\beta \ll \pi^2$.

To conclude, conformal transformations \cite{KG08,RRW09} is a
powerful tool to consider pseudodiffusive transport in undoped
graphene flakes of arbitrary shape, not only without magnetic
field but also in the presence of magnetic fluxes in the system.
Experimental study of the Aharonov-Bohm oscillations and
comparison with simple expressions (\ref{fine1}), (\ref{fine2})
derived here would be a suitable way to check whether the
ballistic regime is reached on not in a given experimental
situation.

\textbf{Acknowledgements.} The work is financially supported by
Stichting voor Fundamenteel Onderzoek der Materie (FOM), the
Netherlands. I am grateful to Pavel Ostrovsky for stimulating
discussions of Ref.\cite{SOGM09} and Carlo Beenakker for helpful
critical remarks.

\end{document}